# Radiation-induced effects in glass windows for optical readout GEM-based detectors

A. Maia Oliveira,[a,b,1] S. Braccini,[b] P. Casolaro,[b] N. Heracleous,[a] J. Leidner,[a] I. Mateu,[b] F. Murtas[a,c] and M. Silari[a]

[a] CERN,
   1211 Geneva 23, Switzerland

[b] Albert Einstein Centre for Fundamental Physics (AEC), Laboratory for High Energy Physics (LHEP),
   University of Bern, Sidlerstrasse 5, 3012, Bern, Switzerland

[c] INFN-LNF,
   00044 Frascati, Italy.

   E-mail: andreia.cristina.maia.oliveira@cern.ch

ABSTRACT: In this paper we present irradiation measurements performed to select a transparent anode substrate that best meets the requirements of an optical readout for a novel detector, the LaGEMPix. The modification of the optical properties of the material due to proton irradiation were studied in soda-lime, fused quartz and fused silica glasses coated with an Indium Tin Oxide layer. The irradiations were performed using the research Beam Transfer Line (BTL) of the IBA Cyclone 18 MeV cyclotron of the Bern University Hospital (Inselspital). We recorded visible scintillation light generated by proton irradiation in the soda-lime and fused quartz samples. We also investigated the darkening of these three glasses and observed radiation-induced colour centres in the soda-lime glass sample. The optical transmission spectra of the samples were measured before and after irradiation. Reductions of 45%, 1% and 0.4% were observed for soda-lime glass, fused quartz and fused silica, respectively (with an associated error of 0.25%). We conclude that the best option for our specific application is the fused silica substrate, which will be the transparent anode for the next generation of the LaGEMPix detector.

KEYWORDS: irradiation, proton beam, radiation damage, soda-lime glass, fused quartz glass, fused silica glass, scintillation, transmittance.

---

[1] Corresponding author.

# Contents

## 1. Introduction

The LaGEMPix is a novel imaging detector with an active area of 60 x 80 mm$^2$ currently under development [1]. It combines a triple Gas Electron Multiplier (GEM) with an area of $10 \times 10$ cm$^2$ continuously flushed with Ar/CF$_4$ gas (90/10 ratio) with a matrix of organic photodiodes (OPDs) coated on an oxide thin film transistor (TFT) backplane produced by Holst Centre/TNO[1]. The detector is designed to be used for quality assurance in hadron therapy, which at present uses protons and carbon ions.

Three GEMs are placed on top of each other to build the triple-GEM as shown in Figure 1. This detector includes a 15 μm thick Mylar window used as the cathode at 3.5 mm from the first GEM, yielding the drift gap between the Mylar cathode and the first GEM. In order to evaluate the feasibility of an optical readout, it is essential to place a transparent electromagnetic shielding below the last GEM and above the readout matrix [2], the ITO (Indium Tin Oxide) coated glass anode shown in Figure 1. This transparent anode collects the drifting electrons while allowing the scintillation photons to pass through, ensuring that photons rather than electrons are responsible for inducing a current in the photodiodes of the image sensor. The probability of a photon reaching the image sensor depends on several factors, including the number of photons that are transmitted by the glass exit window. A higher transmission of the window will translate into a higher induced signal in the readout system. Radiation effects can reduce the window transmission efficiency and radiation damage on specific irradiated areas may induce a non-uniform response. Hence, the selection of the material of the transparent anode is crucial for this application because it affects the response uniformity and the signal stability of the LaGEMPix detector.

---

[1] Holst Centre, PO BOX 8550, 5605 KN Eindhoven, The Netherlands

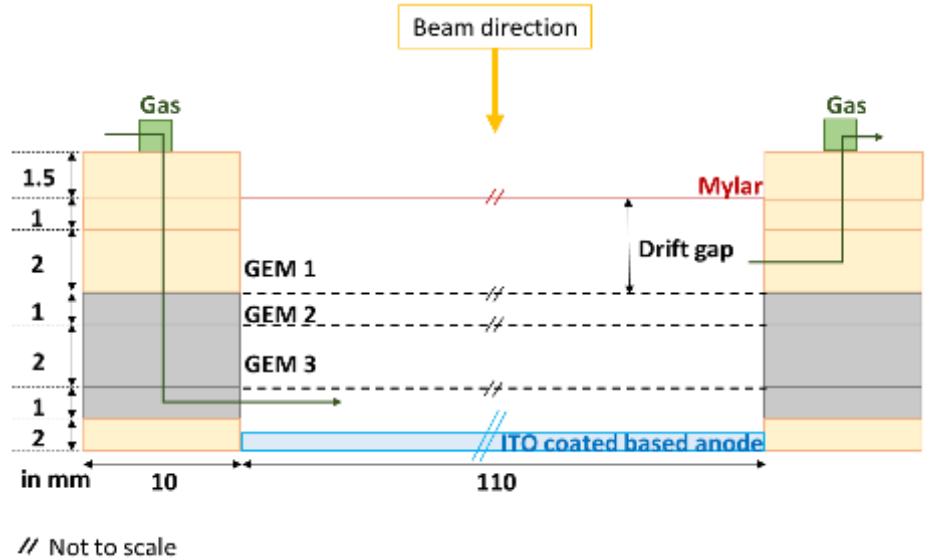

**Figure 1.** Transversal schematics of 10 × 10 cm² triple-GEM assembled to the ITO glass. The optical readout with an active area of 60 × 80 mm² is then placed below the ITO coated based transparent anode.

Optical readout GEM-based detectors coupled to CCD/CMOS cameras used for particle therapy have been studied by several groups [2, 3, 4, 5, 6]. The setups included transparent exit windows, which were in some cases [2, 6, 7] coated with ITO. ITO is used to produce optically transparent anodes due to the combination of electrical conductivity and high light transmission in the visible range. With the exception of Seravalli et al. – who used Duran 50 glass – no decrease in the transmission of the exit window was reported. Scintillation in the exit window was mentioned as a possible source of light production contributing to the noise by Seravalli et al. and Klyachko et al., where it was either regarded as negligible [5] or easy to subtract [4].

We were interested in studying the optical transmission at the visible wavelengths since the Ar/CF$_4$ has a strong visible emission band around 630 nm, which is also well suited for standard optical readouts. Ar/CF$_4$ also presents a broad emission band in the ultra-violet (250 nm) and sharp emission lines above 695 nm [8]. The goal was to determine the best ITO-coated transparent anode material that could be coupled to the triple-GEM. It is important to maximize the transmission of Ar/CF$_4$ scintillation photons to maximize their collection and, consequently, the signal in the readout system.

**2. ITO-based glass materials**

Three different ITO-based anodes were manufactured with the same rounded square shape of 11 × 11 cm² with rounded corners and a thickness of 1.1 mm. Each sample has one surface coated with ITO with a resistivity of 100 Ω/sq. The selected materials are listed in Table 1.

**Table 1.** List of ITO-bases glass materials.

| Material | Provider |
| --- | --- |
| Soda-lime glass | VisionTek Systems Ltd[2]. |

---

[2] VisionTek Systems Ltd., 1 The Acorns, Upton Chester, Cheshire CH2 1JL, United Kingdom, Web: www.visionteksystems.co.uk/

| Heraeus[3] TSC-3 Quartz | Newcastle Optical Engineering Ltd[4] |
| Heraeus Spectrosil® 2000 | Newcastle Optical Engineering Ltd |

The level of purity is vital for high transmission and low radiation damage. Defects like inclusions or impurities present in the glass can originate absorption bands that will reduce the efficiency of the transmission. The manufacturing process affects the quality of the glass and consequently the number of defects. The soda-lime glass is the least pure option and the fused silica the most pure option and therefore the most expensive one.

Duran 50 glass was used as glass exit window in optically read out GEM-based detectors by other groups. The transmission of the exit window decreased in the range 400-600 nm, after 3600 Gy of proton irradiation [9]. This option was therefore excluded in this study.

### 2.1 Soda-lime

The first sample is an ITOGLASS 100P [10] consisting of a polished soda-lime glass with a thickness of 1.1 mm. The glass was coated with a first layer of Silicon Oxide ($SiO_2$) of 25 nm thickness and a second layer of 80 nm of ITO, with a resistivity of 100 Ω/sq.

The exposure to radiation is known to alter the colour of a glass. Depending on the type of glass, the colour induced by radiation can be brown, purple or red [11]. The radiation induced colour centres for the soda-lime were reported by several studies after irradiations with protons, electrons and photons [12, 13, 14]. The non-bridging oxygen hole centres (NBOHCs), which are associated with absorption bands at 623 nm and 412 nm, are the defects responsible for the brown colour after irradiation. The band of 297 nm is recognised as absorption of trapped electrons (TEs) [15].

### 2.2 Fused quartz

According to Heraeus, the manufacturing process regularly used for their *TSC* samples, in particular for Heraeus *TSC*-3 Quartz, is the flame fusion process. This process ensures a high purity material with the lowest possible amount of bubbles and inclusions that are normally present in fused quartz materials [16]. The level of purity is vital for high transmission and radiation insensitivity and therefore the flame fusion process is critical for the quality of the final material.

Mitchell and Paige measured the optical absorption of neutron irradiated fused quartz and reported two absorption bands associated with atomic displacements (oxygen ion vacancies and interstitial oxygen ions) at 163 nm and at 218 nm [17]. The latter could affect the detection of the broad emission band in the ultra-violet of $Ar/CF_4$ gas.

The ATLAS experiment at the CERN LHC [18] tested fused quartz GE214 in long-term operations during proton-proton or heavy ion runs in the Zero Degree Calorimeter and concluded that the specific type of fused quartz is not suitable for extreme radiation environments [19]. However, their measurements were performed for doses of the order of $10^4$ Gy in 10 seconds while we expect doses of the order of 2 Gy per 60 seconds in a clinical hadron therapy beam. Therefore, we included quartz glass in this study.

---

[3] Heraeus Holding GmbH Heraeusstraße, 12-14 D-63450, Hanau Germany, Web: https://www.heraeus.com/

[4] Newcastle Optical Engineering Ltd., Unit 1B, Buddle Industrial Estate, Benton Way, Wallsend, Tyne & Wear NE28 6DL, United Kingdom, Web: www.newcastleoptical.co.uk

**2.3 Synthetic fused silica**

Heraeus Spectrosil® 2000 [20] is a synthetic fused silica glass, which does not contain chlorine, bubbles or inclusions. According to Heraeus, it is an ultra-high purity glass with excellent optical transmission in the deep ultraviolet and visible ranges.

The ATLAS experiment also tested Spectrosil® materials and at 630 nm (broad emission range of Ar/$CF_4$, the selected gas), they measured about 25% relative absorption (irradiated/control). There is an absorption centre at 629 nm due to a NBOHC [21]. The ATLAS and CMS experiments selected fused silica based on its remarkable radiation hardness when compared to other glasses [19, 22].

Other groups have also identified defects in high-purity silica glasses such as oxygen-vacancies and oxygen-interstitial pairs [23, 24]. Cohen and Janezic observed the development of colour centre absorption bands with peaks centred at 629 nm, 463 nm, 310 nm, and 240 nm after X-ray exposure [25].

## 3. Proton beam irradiation

The glasses were irradiated with 17.5 MeV protons at beam intensities of up to 11 nA using the research Beam Transfer Line (BTL) of the IBA Cyclone 18 MeV cyclotron in operation at the Bern University Hospital (Inselspital) [26]. The dose rate (to silicon) on the surface of glass samples was 53.8 Gy/s. A UniBEaM detector [27] was installed in the BTL, as shown in Figure 2. It was used to measure and control the beam characteristics on-line during the irradiations. The UniBEaM measures the beam profiles in both transverse directions by passing scintillating fibres through the beam. The protons are extracted into air by means of a 50 µm stainless steel window, which faces the glass sample. An aluminium disk is placed behind the sample, connected to a Keysight B2985A electrometer to measure the beam current [28]. Each transparent anode was placed perpendicular to the proton beam and with the coated side facing the exit window, which is the same orientation in which they would be placed coupled to the triple-GEM detector.

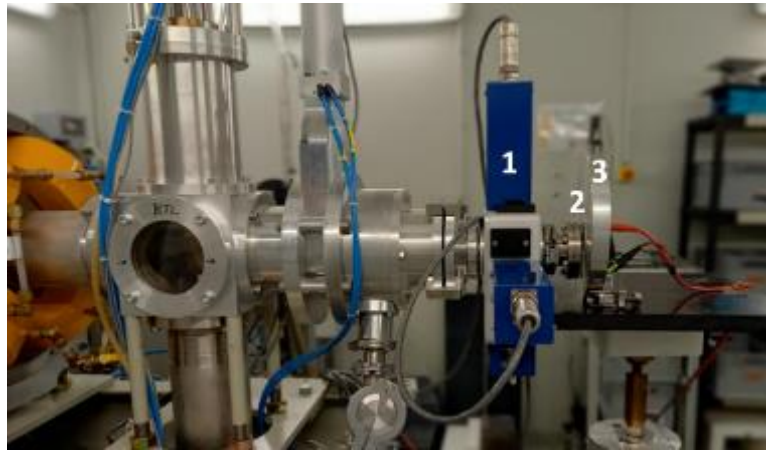

**Figure 2.** The UniBEaM (1) detector and the glass sample (2) installed at the end of the BTL of the IBA Cyclone 18 MeV cyclotron at the Bern University Hospital (Inselspital). An aluminium disk (3) is placed behind the glass sample.

Before irradiating the samples, an EBT3 Gafchromic™ film [29] was positioned in front of the aluminium disk to evaluate the shape and the diameter of the beam. The goal was to have a proton flux similar to a hadron therapy beam. At the National Centre for Oncological Hadron therapy (CNAO) in Italy [30], the proton flux is of the order of magnitude of $10^8$ protons/s/mm$^2$. For a uniform beam of 3 cm diameter from the Bern cyclotron (see Figure 3), it is necessary to set a beam current of 11 nA to reach the same flux.

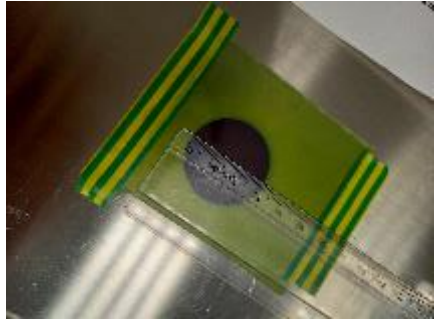

**Figure 3.** Irradiated EBT3 Gafchromic™ film showing a beam spot of 3 cm in diameter.

The horizontal and vertical beam profiles measured by the UniBEaM just before the extraction window are shown in Figure 4. On the basis of the measured beam profile, the average proton flux in the region of interest was evaluated to be $1.5 \times 10^8$ protons/s/mm$^2$.

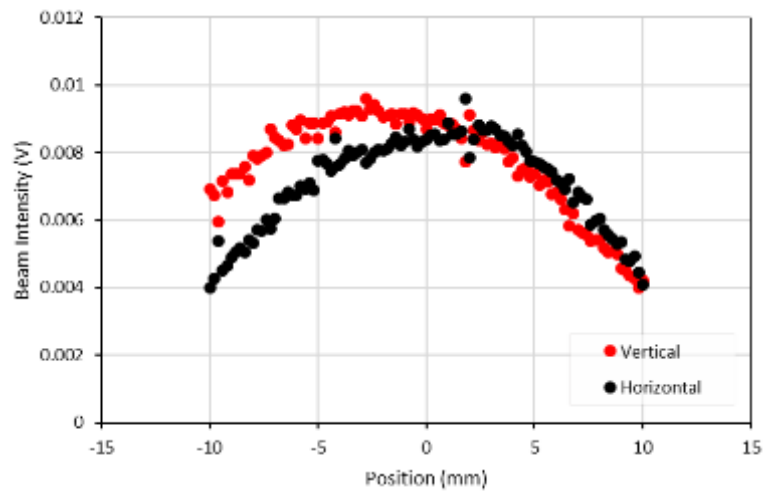

**Figure 4.** Proton beam profiles measured with the UniBEam.

All samples were irradiated two times: a short irradiation with a duration of about 20 seconds to evaluate the scintillation light with a Bosh VTC-206 Mini Bullet Camera, and a 15-minute irradiation to study the optical effects. We removed the aluminium disk shown in Figure 2 during the short irradiation to detect the light with the camera. We placed a camera pointing to the transparent anode in order to record images and to detect the scintillation light emitted by each sample.

An inspection by eye was performed after each irradiation to evaluate the presence of colour centres on each sample.

The radiation-induced effects were evaluated by measuring the transmission of each sample before the irradiation and one week after. The measurements were performed with a Perkin Elmer Lambda 650 UV/VIS Spectrometer (Standard detector Module) in the spectral range from 200 nm to 800 nm, in steps of 1 nm. An intrinsic calibration of the equipment was performed before the measurements. An additional reference measurement with no sample was also carried out to evaluate the fluctuations around 100% level of transmission and to estimate the associated uncertainty of the measurements taken on that day. This uncertainty was estimated by the difference between the maximum value registered at the reference measurements and the 100% transmission level. The samples were placed carefully such that the entire point of light emitted by the spectrometer is within the irradiated area of the glass.

All laboratory experiments were conducted in ambient conditions of temperature, humidity and pressure.

## 4. Results and discussions

### 4.1 Scintillation

The camera detected scintillation generated by proton irradiation in the soda-lime and fused quartz glasses, is clearly visible as bright spots. Figure 5 shows the scintillation light from the fused quartz glass. The mean intensity of the bright spot was analyzed using the ImageJ software [31] by comparing the brightness in images taken during irradiation to those taken without irradiation for a defined rectangular ROI (region of interest). This ROI, which is shown in Figure 5 b), was selected in the centre of the glass sample and it includes the complete bright spot. A relative increase of 66% was measured for soda-lime, 83% for fused quartz and 0.9% for fused silica, which is consistent with the visual observations. The camera did not detect any visible light from the fused silica glass.

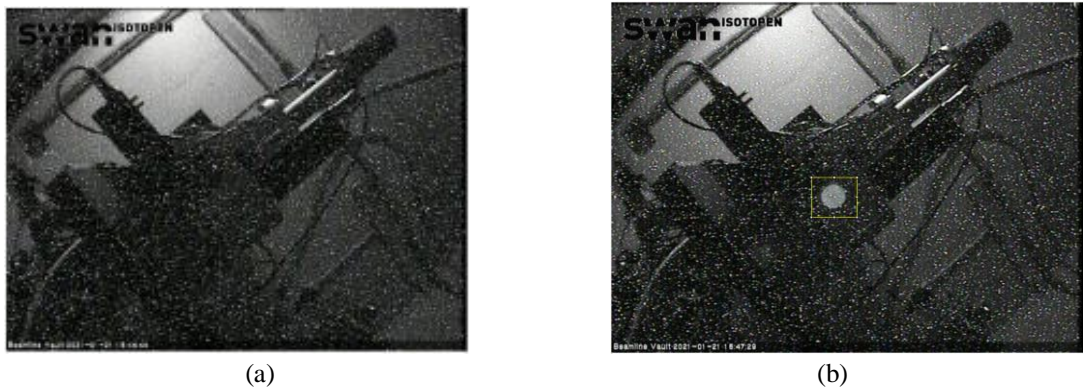

(a) (b)

**Figure 5.** Image of the fused quartz sample when the beam is OFF (a) and when the beam is ON (b). The yellow square determines the ROI selected of the recorded visible light with the ImageJ software.

### 4.2 Transmission

Figure 6 shows the measured transmission spectra of the soda-lime glass sample before and after irradiation. A general degradation in the transmission is observed in all wavelengths in the visible region, as well as an intense reduction centred at the three expected absorption bands. The induced defects of NBOHCs associated with bands at 623 nm and 412 nm are responsible for the brown colour (see Section 4.4). A maximum decrease of 45% in the transmission was measured with an uncertainty of 0.25%.

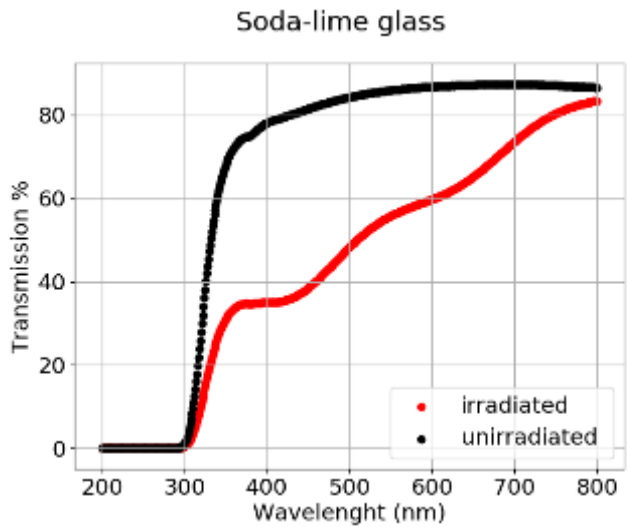

**Figure 6.** Optical transmission spectra of soda-lime glass sample before and after proton beam irradiation. The experimental uncertainties are smaller than the data points.

As all light that is not transmitted must be absorbed, neglecting reflection, the difference $100\% - Transmission\%$ is equal to the Absorption %. Figure 7 shows the absorption as a function of the energy in order to compare directly the centres of the three emission bands with the results reported by A. Serrano et al. [11]. The centres of the peaks' absorption bands, which corresponds to the mean values of each Gaussian, were obtained by fitting the profile by a triple Gaussian function. Table 2 shows that the present results are in agreement with literature data within maximum 2.2%.

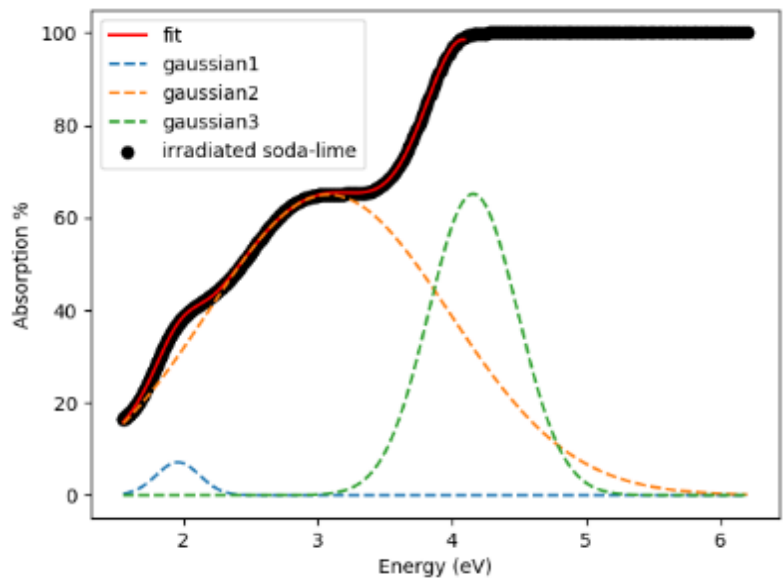

**Figure 7.** Optical absorption spectrum of proton irradiated soda-lime glass.

**Table 2.** Centres of the peaks of the three Gaussian fits to the optical absorption spectrum. The present results are compared with those from ref. [15].

| Energy (eV) [11] | Energy (eV) [this work] | % difference |
|---|---|---|
| 1.990 ± 0.001 | 1.958 ± 0.001 | 1.6 |
| 3.012 ± 0.002 | 3.078 ± 0.002 | 2.2 |
| 4.171 ± 0.001 | 4.157 ± 0.003 | 0.3 |

Figure 8 shows the measured transmission spectra of the fused quartz glass sample before and after irradiation. A maximum decrease of 1% was measured with an uncertainty of 0.25%.

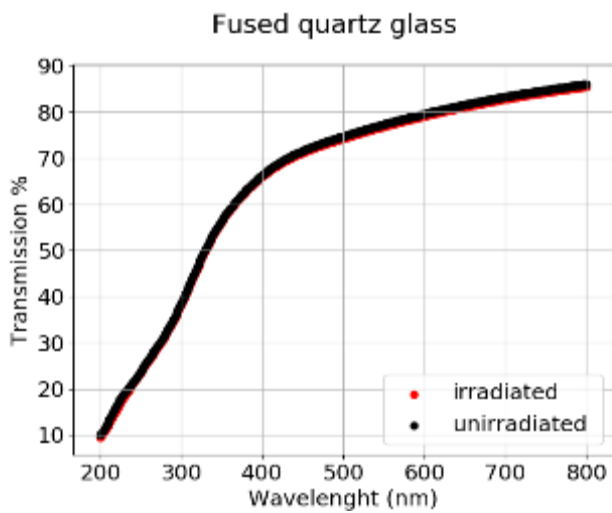

**Figure 8.** Optical transmission spectra of fused quartz glass before and after proton beam irradiation. The experimental uncertainties are smaller than the data points.

Figure 9 shows the measured transmission spectra of the fused silica glass sample before and after irradiation. A maximum decrease of 0.4% was measured with an uncertainty of 0.25%.

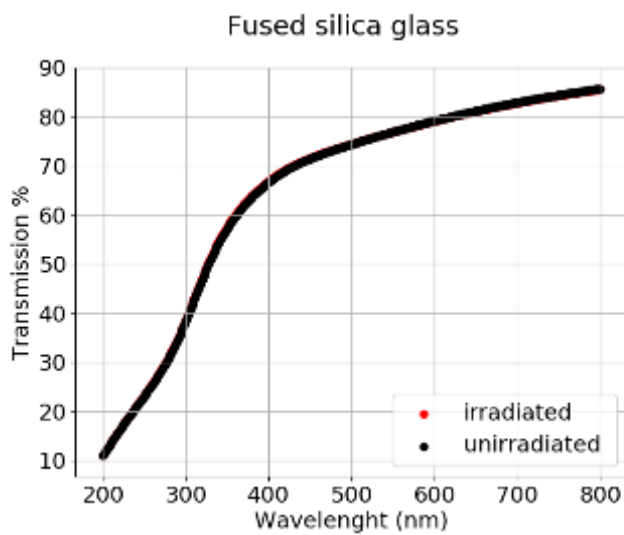

**Figure 9.** Optical transmission spectra of fused silica glass sample before and after proton beam irradiation. The experimental uncertainties are smaller than the data points.

The fused silica substrate presents the smallest transmission loss. Consequently, a more consistent number of photons will reach the readout over time for the fused silica glass when compared with other samples.

**4.3 Sheet resistivity**

To estimate the resistivity of each type of glass, we produced a 19-pin probe to obtain a more accurate Ω/sq value rather than using a standard 2-pin multi-meter [32], which only gives a point-to-point measurement. Each sample was divided into small square regions and the resistivity was measured before and after the irradiations. For the soda-lime glass, values from 80 to 93 Ω/sq were measured, for the fused quartz values from 160 to 480 Ω/sq and fluctuations between 120 and 240 Ω/sq were observed for fused silica. The differences between the resistivity of soda-lime sample and the other two samples may be due to the fact that the samples are produced by different manufacturers. No alteration in the sheet resistivity of the samples was registered due to irradiation. Wei et al. [33] reported an alteration in the optical properties of ITO films deposited on quartz substrate after irradiation with 100 keV protons. They measured an increase of the sheet resistivity using the four-point probe method when the fluence exceeded $2\times10^{16}$ cm$^{-2}$.

**4.4 Colour centres**

To evaluate the quick formation of colour centres, it is necessary to compare the pictures of the ITO soda-lime glass sample for different irradiation periods. After the first irradiation lasting 20 seconds, a brownish circle was observed, while after the second irradiation of 15 minutes the circle became darker, as depicted in Figure 10.

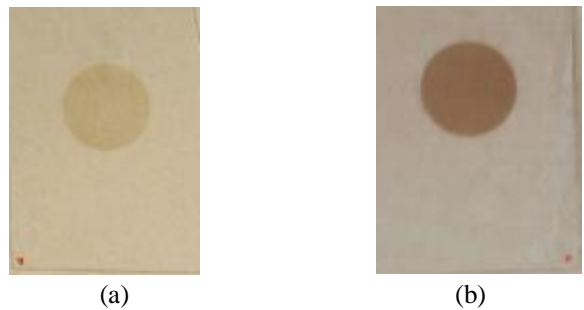

(a)  (b)

**Figure 10.** Visible radiation damages of ITO soda-lime glass after (a) short - 20 seconds - and (b) long -15 minutes - irradiations.

The opacity caused by the radiation in ITO soda-lime glass is clearly visible by eye. The colour of the darkened circle after 15 minutes and 20 seconds irradiations was measured using a colorimeter Pantone Capsure™ Model RM200 [34]. This colorimeter allows the identification of colours according to Pantone colour libraries. The colour Pantone® 452C has been identified in the darkened circle that corresponds to Red=176, Green=170, Blue=126 in the RGB system, one of most common ways to encode colours. The device was incapable of identifying the precise colour in transparent regions, i.e. the non-irradiated area in the soda-lime glass and the whole area of the other two samples (fused quartz and fused silica). Similar effects on this type of substrate under proton irradiation have been reported in the literature [7]. There are radiation-induced absorption sites in the glass, and the number of these sites increases with the number of impurities present in the glass.

After radiation exposure, the soda-lime glass shows a drastic reduction in the transmission. One common method to recover transparency is to heat the glass. However, after the transparent anode has been assembled in the LaGEMPix detector, it is not possible to submit the full detector to an annealing above 100 °C to allow the recovery of these absorption bands. Therefore, soda-lime is not a viable option for the substrate of the anode and a more radiation-tolerant material must be selected. On the other hand, no creation of colour centres was observed on fused quartz and fused silica samples after both irradiations.

### 4.5 Summary of results

Table 3 summarizes the experimental results. Visible scintillation light generated by proton irradiation was observed in the soda-lime and fused quartz samples. The darkening of the three materials was also qualitatively investigated; radiation-induced colour centres in the soda-lime glass sample were observed. The optical transmission spectra of the three samples were measured after irradiation. A decrease in the optical transmission of 45%, 1% and 0.4% was measured for soda-lime glass, fused quartz and fused silica respectively (with an associated error of 0.25%).

**Table 3.** Summary of the radiation-induced effects for the three tested glasses.

| Glass | Scintillation | Irradiation induced colour centres | Reduction in transmission (maximum difference) | Variations in the sheet resistivity |
|---|---|---|---|---|
| Soda-lime | Yes – 66% | Yes (R=176, G=170, B=126) | 45 % | Not observed |
| Fused quartz | Yes – 83% | Not observed | 1% | Not observed |
| Fused silica | Not observed – 0.9 % | Not observed | 0.4% | Not observed |

### 5. Conclusions

The soda-lime glass is not a suitable option for the optical readout under study, since the loss in transmission would affect the light signal reproducibility of the LaGEMPix detector. Fused quartz appears to be more tolerant to radiation. However, it scintillates, which will add a signal to the scintillation photons emitted in the triple-GEM. Based on the measurements, the best option is the fused silica substrate, since it presents the lowest transmission loss. The incorporation of this glass as exit window will produce a higher and more reliable signal over time when compared with the other materials investigated. The results here reported motivated a replacement of the transparent anode material for the next generation of the LaGEMPix detector for applications in hadron therapy.

### Acknowledgments

The authors wish to thank Rui de Oliveira for the materials, and Thomas Schneider for the technical help and the measurements performed at the Optical Quality Control Lab - Thin Film & Glass service of CERN, and Lucia Gallego for providing the image that was adapted to Figure 1, and Imprimerie Villière – Villir (Beaumont, France) for providing the colorimeter Pantone Capsure™ Model RM200. This Project has been co-funded by the CERN Budget for Knowledge Transfer to Medical Applications. This project has also received funding from the ATTRACT project funded by the EC under Grant Agreement 777222. Andreia Maia Oliveira was co-supported by a grant from FCT (Portugal) with reference SFRH/BEST/142965/2018. This research was partially funded by the Swiss National Science Foundation (SNSF) grant CRSII5_180352.